# Anomalous Hall Effect and Magnetoresistance of SrFe$_{1-x}$Co$_x$O$_{3-\delta}$

Takahiko Ido, Yukio Yasui and Masatoshi Sato

*Department of Physics, Division of Material Science, Nagoya University, Furo-cho, Chikusa-ku, Nagoya 464-8602*

**Abstract**

Transport and magnetic studies on polycrystalline samples of SrFe$_{1-x}$Co$_x$O$_{3-\delta}$ have been carried out to investigate the relationship between the magnetic structure and the anomalous Hall resistivity $\rho_H$. The hysteretic behavior of the magnetization observed in the measurements with varying temperature $T$ up and then down after zero field cooling indicates that the system has the reentrant spin-glass phase, which is supported by the increasing width of the magnetic reflections observed by neutron diffraction with decreasing $T$ below the Curie temperature $T_C$. Detailed analyses of the observed Hall resistivity $\rho_H$ indicate that the anomalous Hall coefficient exhibits unusual behavior in the reentrant spin-glass phase. The magnetic field ($H$)- and $T$-dependence of the magnetoresistance of the present system can be understood by a spin dependent tunneling model.

corresponding author : M. Sato (e-mail : msato@b-lab.phys.nagoya-u.ac.jp)

## Introduction

SrFe$_{1-x}$Co$_x$O$_{3-\delta}$ with perovskite structure exhibits the helical spin ordering in the region of $x<0.2$,[1] where the propagation vector $\boldsymbol{q}$ is along the [111] direction and has the absolute value of ~$0.112|\boldsymbol{a}^*|$[1] with $\boldsymbol{a}^*$ being the reciprocal lattice unit vector. For $x>0.2$, the system has the ferromagnetic moment.[2-6] The electrical resistivity ρ decreases with increasing $x$ and also depends on the amount of the oxygen deficiency δ.[2,7] The metallic state can be realized by synthesizing samples under the high oxygen pressure[3] or by synthesizing them by an electrochemical oxidation method.[8]

We expected the non-trivial spin structure even in the ferromagnetic region of $x$ of SrFe$_{1-x}$Co$_x$O$_{3-\delta}$ and adopted it as one of candidate systems for which the relationship between the non-trivial magnetic structure and the transport properties could be studied. In particular, we pay attention to the relationship between the non-trivial magnetic structure and the anomalous Hall resistivity, because the problem seems not to be fully understood by existing classical theories. For example, the magnetic field($H$)- and temperature($T$)-dependence of $\rho_H$ of the pyrochlore molybdate Nd$_2$Mo$_2$O$_7$ with non-trivial magnetic structure[9,10] cannot be well understood.[10-14] In the studies of this problem, a possible role of the spin chiral order has been proposed.[14,15] Although the present authors' group has pointed out that the idea cannot describe the main behavior of the Hall resistivity $\rho_H$ of Nd$_2$Mo$_2$O$_7$, it is interesting to search systems in which the ordering of the spin chirality χ explicitly contributes to $\rho_H$.[16] From this point of view, it is also interesting, as pointed out by Tatara and Kawamura[17] to investigate the behavior of $\rho_H$ in the spin glass phase, where the spin chirality is considered to be in the frozen state.

As we expected, the magnetization $M$ measured for polycrystalline samples of SrFe$_{1-x}$Co$_x$O$_{3-\delta}$ ($0.0 \leq x \leq 0.8$) does not saturate up to the magnetic field ($H$) of 5 T at 5 K, indicating the magnetic structure is non-trivial even in the ferromagnetic region of $x$. In the present paper, results of the magnetic and transport measurements carried out on this system are mainly reported. We have found that the system has the reentrant spin glass state and that $\rho_H$ exhibits unusual behavior in the spin glass phase. It has been also found that the magnetoresistance can be understood by a model of metallic grains connected with insulating barriers, where the barrier height for the tunneling electrons depends on their spin directions.

## Experiments

Polycrystalline samples of SrFe$_{1-x}$Co$_x$O$_{3-\delta}$ were synthesized from the mixtures of SrCO$_3$, α-Fe$_2$O$_3$ and Co$_3$O$_4$ with nominal molar ratios, where all these initial materials were 99.9 % pure: The mixtures were ground, pressed into pellets and heated for about 24 h at 1100 °C. Then, the samples were furnace cooled to room temperature. They were ground and pelletized again and the same heat treatment was carried out. After these heat treatments, the pellets were annealed at 320 °C under the oxygen pressure 60 atm for 2 days in order to reduce the amount of the oxygen deficiency $\delta$.

The lattice constant $a$ was determined by X-ray diffraction. The oxygen deficiency $\delta$ was estimated by the thermogravimetry(TG) carried out in the flowing $N_2$ gas. The electrical resistivities $\rho$ were measured by the conventional four probe method. The magnetizations $M$ were measured by using a SQUID magnetometer: The $M$-$T$ curves were taken after the zero field cooling with increasing and then decreasing $T$ under the field of $H$=1 T. The $M$-$H$ curves were taken with increasing $H$ stepwise after the zero field cooling at various fixed temperatures. The Hall resistivity $\rho_H$ was measured after the zero field cooling by rotating the samples with respect to the field direction and then increasing $H$ stepwise. The same sample was used for the measurements of $M$-$H$ and $\rho_H$-$H$ curves and the field was always applied perpendicular to the sample plate in order to avoid the ambiguities which may arise from the difference between the demagnetization fields in the two kinds of measurements. The magnetoresistance $\Delta\rho$ was measured after the zero field cooling by the four probe method, where **H** was applied perpendicular to the current direction and increased stepwise at various fixed temperatures. Neutron diffraction study was carried out for $x$=0.5, $\delta$=0.14 with the T1-1 triple axis spectrometer at JRR-3M at Tokai. The neutron wavelength was ~2.459 Å.

**Experimental results**

Figure 1 shows the $x$-dependence of the lattice constant $a$ and the oxygen deficiency $\delta$. The lattice constant $a$ exhibits the $x$-dependence slightly different from that of the previous study,[3,18] which can be attributed to the different oxygen annealing conditions. The $x$-dependence of $\delta$ can be understood by considering the difficulty to stabilize Co(IV).

Figure 2 shows the $T$ dependence of the resistivity $\rho$ for the polycrystalline samples with various $x$ values. For $x$<0.2, $\rho$ exhibits rather rapid increase with decreasing $T$, while the $\rho$ for $x$≥0.2 is nearly independent of $T$ above ~60 K and slightly increases in the low $T$ region.

$M$-$T$ curves taken with $H$=1 T for various $x$ values are shown in Fig. 3. Ferromagnetic or nearly ferromagnetic behavior can be seen for $x$≥0.2. The Curie temperature $T_c$ roughly estimated by using the Arrott plot ($M^2$-$H$/$M$ plot) varies almost linearly in $x$ from ~80 K at $x$=0.5 to ~160 K at $x$=0.8. ( The $T_c$ values are consistent with those determined as the inflection points of the $M$-$T$ curve taken with $H$=100 Oe.) For the region of $0.2 \leq x \leq 0.4$, it is difficult to distinguish if there exists the long range ferromagnetic ordering at finite $T$ in the zero magnetic field. (Even for such the $x$ region, the large uniform magnetization $M$ exists in the measurement of $\rho_H$, because it is induced by the applied fields.) The magnetic moments observed at 5 K and the Curie temperatures $T_C$ are smaller than those reported in the previous paper,[3] probably because of the larger oxygen deficiency of the present samples. The hysteretic behavior can be observed below $T \equiv T_g^* \sim 60$ K for all samples shown in Fig. 3 and it is more significant below $T \equiv T_g \sim 30$ K for $0.2 \leq x \leq 0.6$.

The $M$-$H$ curves measured up to $H = 5$ T in the wide $T$-range are shown in Fig. 4 for $x = 0.3$ (top), 0.5 (middle) and 0.7 (bottom), for example. $M$ does not saturate up to the maximum field used here (5 T), and in the relatively low field region ($H \leq 1$ T), $M$ decreases with decreasing $T$ below ~70 K. In order to clarify origin(s) of these behaviors of $M$-$H$ curves, neutron diffraction experiment for the sample with $x = 0.5$ and $\delta = 0.14$ was carried out, where magnetic reflections were observed only around the nuclear Bragg points. (No superlattice reflection, which indicates the existence of the long range order of modulated spin structure, has been observed.) The intensity profiles of the reflections around the nuclear Bragg points were obtained by subtracting the intensities in the paramagnetic phase. In Fig. 5(a), the resulting profile of the magnetic reflection around the (1,0,0) point is shown, for example, and the profile width (full width at half maximum) is shown in Fig. 5(b) as a function of $T$. The width increases with decreasing $T$, indicating that the linear size of the ferromagnetic domain, $L_f$ decreases with lowering $T$. Because the $T$-dependence of the profile width or that of $L_f$ becomes significant along with the increasing hysteretic behavior of the $M$-$T$ curve observed below $T_g^*$ and $T_g$, the behavior can be considered to be due to the occurrence of the reentrant spin glass phase, and $T_g$ and $T_g^*$ defined above can be considered as the characteristic temperatures of the freezing transition.

To estimate the $L_f$ value for $x=0.5$ at 10 K from the profile in Fig. 5(a), we have to clarify if the profile contain a Gaussian component corresponding to the existence the uniform magnetization. However, a trial to divide the profile into two components, Gaussian and Lorentzian ones has turned out not to be very successful because of the insufficient statistics of the data obtained by subtracting the nuclear scattering contribution for the present powder sample. Then, we can just estimate the upper bound of $L_f$ to be ~200 Å at 10 K, as the value deduced by assuming that the Gaussian component does not exist. (If the Gaussian component or the corresponding uniform magnetization exists, $L_f$ may become as small as ~100 Å at 10 K.)

In Fig. 6, the Hall resistivities $\rho_H$ are shown for various $x$ values at several fixed temperatures. The data in the $H$-region between 2 T and 5 T are fitted by the ordinary expression $\rho_H = R_0 H + 4\pi R_s M$, where $R_0$ is the ordinary Hall coefficient and $R_s$ is the anomalous Hall coefficient. (Strictly speaking, the equation can be used for thin plate-like samples in which the demagnetization field is, as in the present case, well approximated to be $-4\pi M$,[9,16]). The results of the fittings are shown by solid lines in the figures for representative temperatures (5 K and 200 K, for example). The lines in the region of $H$<2 T are calculated by using the observed $M$ for the parameters obtained by the fittings. We find significant deviation of the calculated curves, in this $H$ region, from the observed data. In Fig. 7, the $T$-dependence of the fitting parameters $R_s$ is shown for various $x$ values. In the relatively high $T$ region(<200 K), the $R_s$ decreases with decreasing $T$ as for ordinary ferromagnets. However, for $0.2 \leq x \leq 0.6$, it increases with decreasing $T$ in the low $T$ region.

Figures 8 (a) and 8(b) show the magnetoresistances $\Delta\rho$ scaled by $\rho_0\{\equiv\rho(H=0)\}$ against $H$ at various fixed $T$ values. For all the $x$ values, the negative magnetoresistance can be seen as reported in the previous papers.[18,19] At relatively high temperatures, the $\Delta\rho/\rho_0$-$H$ curves are concave and at low temperatures, the curves exhibit the concave-convex crossover behavior with increasing $x$.

### Discussion

Considering that the hysteretic behavior is observed between ZFC and FC in the $M$-$T$ curves and that the profile width of the neutron magnetic reflection observed around nuclear Bragg points increases with decreasing $T$, it can be said that the reentrant spin glass state appears in the relatively low field at low temperatures. The glassy behavior is most significant for the samples of $0.2\leq x\leq 0.6$. The local spin structure may be non-trivial, because $M$ does not saturate up to the magnetic field as high as 5 T, even though superlattice reflections which indicate the existence of the (quasi) long range order of modulated spin structure, have not been observed. We speculate that the spin structure is derived by superposing a uniform (ferromagnetic) spin component on the helical structure which is similar to the one reported for $x = 0.0$.[1]

In the $T$-region where the hysteretic behavior of the $M$-$T$ curves in Fig. 3 is not significant, the $H$-dependence of the observed Hall resistivities $\rho_H$ can be fitted rather well by the equation $\rho_H = R_0 H + 4\pi R_s M$, which is usually used for ordinary ferromagnets. In contrast, in the low $T$ region where the hysteretic behavior is significant, we cannot obtain satisfactory fits. One might think that this deviation of the $\rho_H$-$H$ curves from the relation $\rho_H=R_0H+4\pi R_sM$ is due to possible difference (or non-reproducibility) between the magnetizations $M$ realized in the $M$- and $\rho_H$-measurements. (The processes of these measurements are already described above.) However, the possibility is ruled out, because the reproducibility of the *M-H* curves has been confirmed to be very well in the measurements, irrespective of measuring procedures with and without rotating samples with respect to *H*. (It can be also said that the time dependence of *M* is negligible even in the glassy phase.)

In the relatively high *T* region, $R_s$ decreases, as for ordinary ferromagnets, with decreasing *T*. However, for the samples of $0.2 \leq x \leq 0.6$, in which the glassy behavior can be seen, it increases with decreasing *T* below ~60 K. To understand this unusual behavior, we may have to consider the relation $R_s \propto \rho^2$ derived by the classical work of Kurplus and Luttinger[20] and attribute the observed upturn of the $R_s$-*T* curves at ~60 K is due to the increase of $\rho$ with decreasing *T*. However, the possibility is ruled out, because if the idea is correct, the $R_s$-*T* curves of the sample with *x*=0.7 and 0.8 should exhibit the upturn at ~60 K, which is not observed. We think that the behavior of $\rho$ is just due to the grain boundary effect which does not contribute to the behavior of $\rho_H$. Then, we can say that the anomalous behavior of the $\rho_H$-*H* and $R_s$-*T* curves has been observed in the *x* region, where the reentrant spin glass phase is significant.

As one of the candidate mechanisms which can explain the deviation observed between the experimental data and the fitted curves of $\rho_H$, the mechanism proposed by Tatara and Kawamura[17] is attractive. They consider the coupling of the uniform magnetization *M* with the spin chirality *c* frozen in the reentrant spin glass phase. The uniform component of *c* induced by the coupling contribute to the Hall resistivity in addition to the components ever known.[20,21]

We add here that the *T*-dependence of $\rho_H$ for a sample with *x*=0.0 is consistent with that reported by Hayashi *et al*.[22] for film samples, where the anomalous decrease of $\rho_H$ was found to appear at $T_N$ with decreasing *T* and then $\rho_H$ exhibits rapid increase with further decreasing *T*.

To explain the behavior of the observed magnetoresistance $\Delta\rho$ or $\Delta\rho/\rho_0$, we propose the spin dependent tunneling model, where metallic grains are separated, as shown in Fig. 9(a), by an insulating potential barrier. We presume that the existence of this kind of inhomogeneity is caused by the disorder of oxygen deficiency and the random distribution of Fe and Co atoms. We consider a case where conduction electrons in the *i*-th metallic grain with the magnetization $M_i$ tunnel to the neighboring *k*-th grain with the magnetization $M_k$ through the *j*-th potential barrier with the magnetization ***M***$_j$. The tunneling probability *t* can be given by

$$t \propto \exp(-2\kappa d) \qquad (1)$$

where $\kappa$ is the inverse penetration depth of the electron wave function and *d* is the width of the barrier. The barrier height $V$ for the conducting electrons depends on their spin directions, because the exchange coupling of the electrons with the magnetization $M_j$ exists. Then, we describe the height by using the magnetizations of the grains as,

$$V = \Delta - \alpha M_i \cdot M_j \qquad (2)$$

In eq. (2) $\Delta$ is the barrier height (measured from the Fermi energy) without the exchange coupling and $\alpha$ is the coupling constant. When the magnetic field is applied, the average value of the $M_i \cdot M_j$ becomes positive and $\kappa$ changes from $\kappa_0$ to $\kappa_0 - \delta\kappa$. The change $\delta\kappa$ is expressed by the relation,

$$\kappa^2 = (\kappa_0 - \delta\kappa)^2 = 2m(\Delta - \alpha M_i \cdot M_j)/\hbar^2 \qquad (3)$$

and

$$\delta\kappa \cong m\alpha(M_i \cdot M_j)/\kappa_0 \hbar^2 \qquad (3'),$$

where $m$ is the electron mass, and we assumed the relation $\kappa_0 >> \delta\kappa$ or $\Delta >> \alpha M_i \cdot M_j$. Then, the electrical resistivity is given by

$$\rho = \rho_0 \cdot \exp[-2\delta\kappa \cdot d] \qquad (4)$$

by using the resistivity $\rho_0$ for $H=0$. Expanding eq. (4) for $2\delta\kappa\, d<<1$, we obtain following equation,

$$\frac{\Delta r}{r_0} \propto \frac{2mda}{k_0 \hbar^2}(2\overline{M}\delta M + \delta M^2), \qquad (5)$$

where $M$ is divided into $\overline{M}$ (spontaneous magnetization) and $\delta M$ (magnetic field induced magnetization). The former term is zero above $T_C$. It should be noted here following things. If the grains have both the spin up($\uparrow$) and spin down($\downarrow$) electrons, the electronic state densities for two spin directions $\uparrow$ and $\downarrow$ at the Fermi surface, $N_\uparrow(\varepsilon_F)$ and $N_\downarrow(\varepsilon_F)$ have to be different to produce the finite magnetoresistance (see Fig. 9(c)), because the effect of the magnetic field $H$ on $\Delta\rho$ through the relation (3') cancels for two spin directions, even if the magnetization directions of all grains are aligned by the field $H$. In the actual separation of the total $M$ into $\delta M$ and $\overline{M}$, we roughly assume that $\delta M$ is proportional to the Brillouin function ($\delta M \propto B_J(gJ\mu_B B/k_B T)$)[23] and $\overline{M}$ is constant in the $H$-region larger than 2 T, because the ferromagnetic moments of all domains are aligned. We have examined if the relation (5) can describe the observed behavior of the magnetoresistance, optimizing $\overline{M}$ and the effective $J$ value.

The $J$ value of the order of 100 found in the analyses implies that the longitudinal(amplitude) moment fluctuation of an effective the magnitude ~100 $\mu_B$, exists in the ferromagnetic domains with the size $L_f$. The average diameter of the effective moments is ~20 Å. It is much smaller than the upper bound of the ferromagnetic domain size $L_f$. Because the magnetoresistance $\Delta\rho$ is determined in the present model by the barriers located between the boundary region of ferromagnetic domains, the metallic regions (denoted by $i$ and $k$ in Fig. 9(a)) may not be much smaller than $L_f$ to produce the observed $\Delta\rho/\rho_0$ (>0.1).

In Fig. 10, we show the results of the analyses for $x$=0.3, 0.5 and 0.8. The magnetoresistance $\Delta\rho/\rho_0$ is nearly linear in ($2\overline{M}\delta M+\delta M^2$) as is predicted in eq. (5) and the proportionality constants $C$ are roughly independent on $T$. These results indicate that the present tunneling model can describe the experimental data well. For the explanation of the observed behavior of the magnetoresistance of (Nd,Sr)MnO$_3$, a spin dependent variable range hopping model was used by Wagner et al. [23] Their model predicts that $C \propto 1/T$, while it is independent of $T$ in the present model, as is actually observed.

In summary, $M$-$H$ curves, neutron diffraction, Hall resistivity and magnetoresistance of SrFe$_{1-x}$Co$_x$O$_{3-\delta}$ ($0.0 \leq x \leq 0.8$) have been studied. The observed deviation of the $\rho_H$-$H$ curves from the relation $\rho_H = R_0 H + 4\pi R_s M$ and the anomalous $T$-dependence of $R_s$ observed in the reentrant spin glass phase indicate that the transition to the glass phase brings about a certain change of the $\rho_H$-behavior. It is tempting to relate it to the effect of the uniform chirality proposed by Tatara and Kawamura.[17] Further investigation is necessary to clarify the role of the chirality in the determination on the Hall resistivity $\rho_H$. To describe the observed behavior of the magnetoresistance, the spin dependent tunneling model has been introduced and rather satisfactory agreements with the observed data have been obtained.

Acknowledgment: The authors thank Professor T. Terashima for sending a preprint on the Hall coefficient of SrFeO$_3$ films. They also thank Professor H. Kawamura for stimulating discussion.

Figure captions

Fig. 1  Lattice parameter $a$ and oxygen deficiency $\delta$ of $SrFe_{1-x}Co_xO_{3-\delta}$ are plotted against $x$.

Fig. 2  $T$-dependence of $\rho$ of $SrFe_{1-x}Co_xO_{3-\delta}$ is shown for various $x$ values.

Fig. 3  Magnetizations $M$ measured with $H$=1 T are plotted against $T$ for various $x$ values. Solid- and dashed-lines are drawn through the data taken by the zero-field- and field-coolings, respectively. The arrows indicate the characteristic temperatures, $T_g^*$ and $T_g$ for $x$=0.5, for example.

Fig. 4  $M$-$H$ curves taken for the $SrFe_{1-x}Co_xO_{3-\delta}$ samples $x$=0.3, 0.5 and 0.7 are shown at various fixed temperatures.

Fig. 5  (a) Profile of the neutron magnetic reflection taken around 100 nuclear Bragg reflection at 10 K for $SrFe_{0.5}Co_{0.5}O_{2.86}$. (b) $T$-dependence of the profile width of the magnetic reflection around 100 nuclear Bragg reflection.

Fig. 6  Hall resistivities $\rho_H$ taken for $x$=0.3, 0.4, 0.5 and 0.7 are shown for examples at various temperatures $T$. Solid lines show the results of the fittings to the data in the $H$ region larger than 2 T taken at 5 K and 200 K, by using the expression $\rho_H = R_0 H + 4\pi R_s M$. The lines in the region of $H$<2 T show the $\rho_H$-$H$ curves calculated by using the observed $M$ for the parameters obtained by the fittings.

Fig. 7  Anomalous Hall coefficients $R_s$ are plotted against $T$ for various $x$ values.

Fig. 8(a)  Magnetoresistance $\Delta\rho/\rho_0$ is plotted against $H$ at various fixed temperatures for $x$=0.2 (top), 0.3 (middle) and 0.4 (bottom).

Fig. 8(b)  Magnetoresistance $\Delta\rho/\rho_0$ is plotted against $H$ at various fixed temperatures for $x$=0.5 (top), 0.6 (middle) and 0.8 (bottom).

Fig. 9  (a) Schematic model used to describe the behavior of the observed magnetoresistance. (b) An example set of $\overline{M}$ (spontaneous magnetization) and $\delta M$ (magnetization induced by $H$) derived by assuming that $\delta M \propto B_J(gJ m_B B/k_B T)$ and $\overline{M}$ is constant in the region of $H$>2 T. (c) Illustration of the spin dependent electron filling.

Fig. 10  ($2\overline{M}\delta M + \delta M^2$) is plotted against $-\Delta\rho/\rho_0$ at various temperatures for $x$=0.3 (top), 0.5 (middle) and 0.8 (bottom).

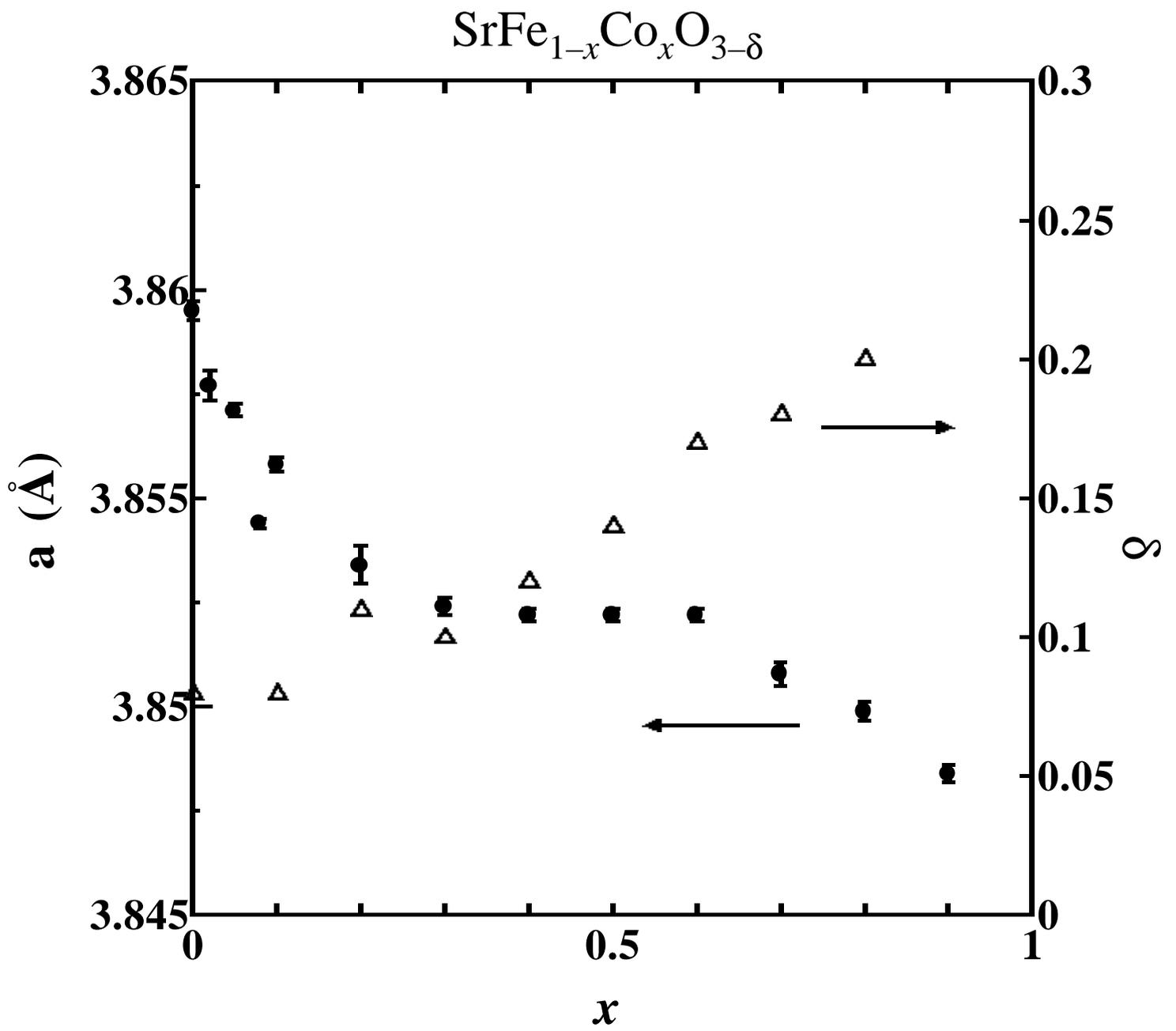

Fig. 1

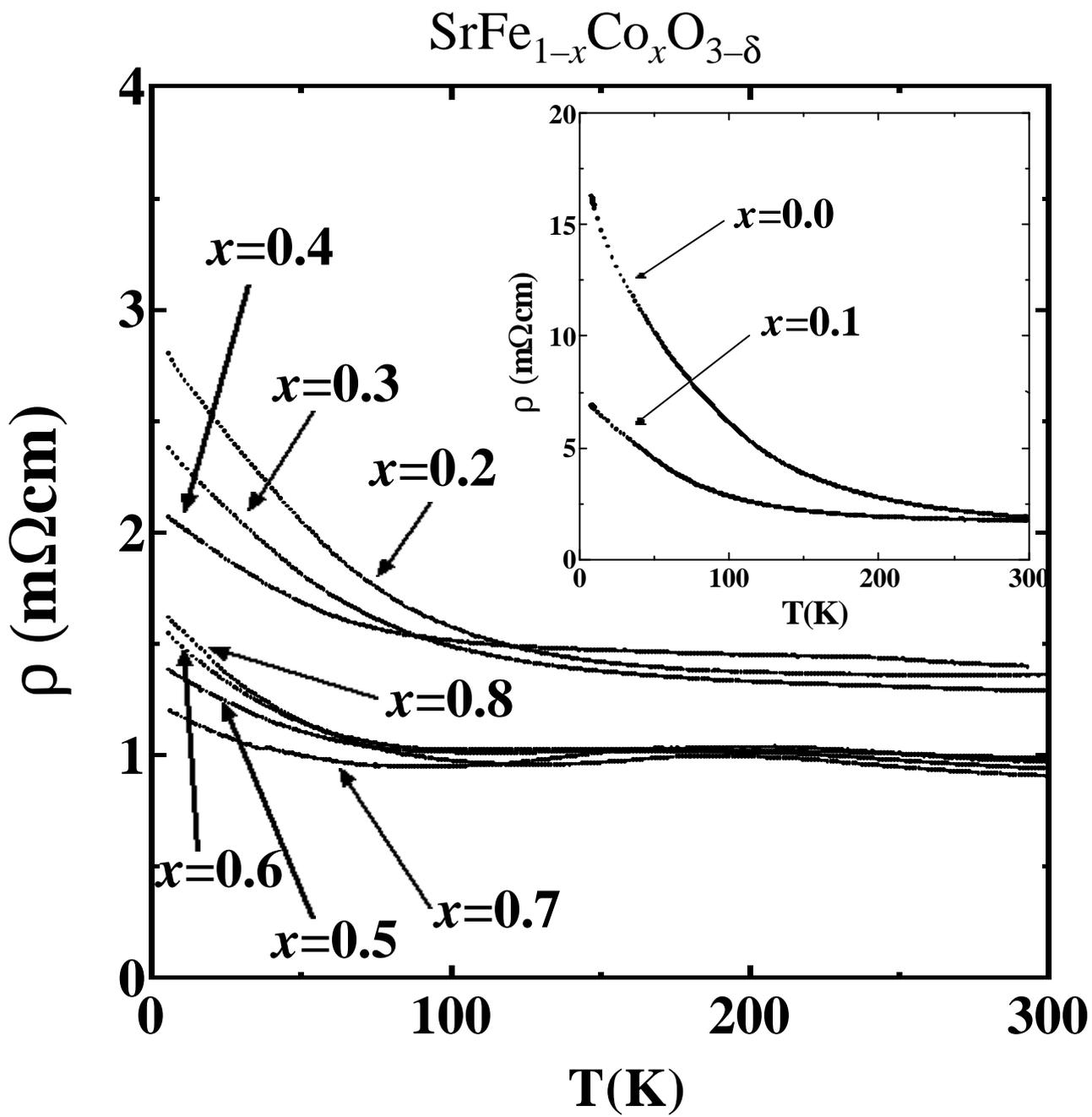

Fig. 2

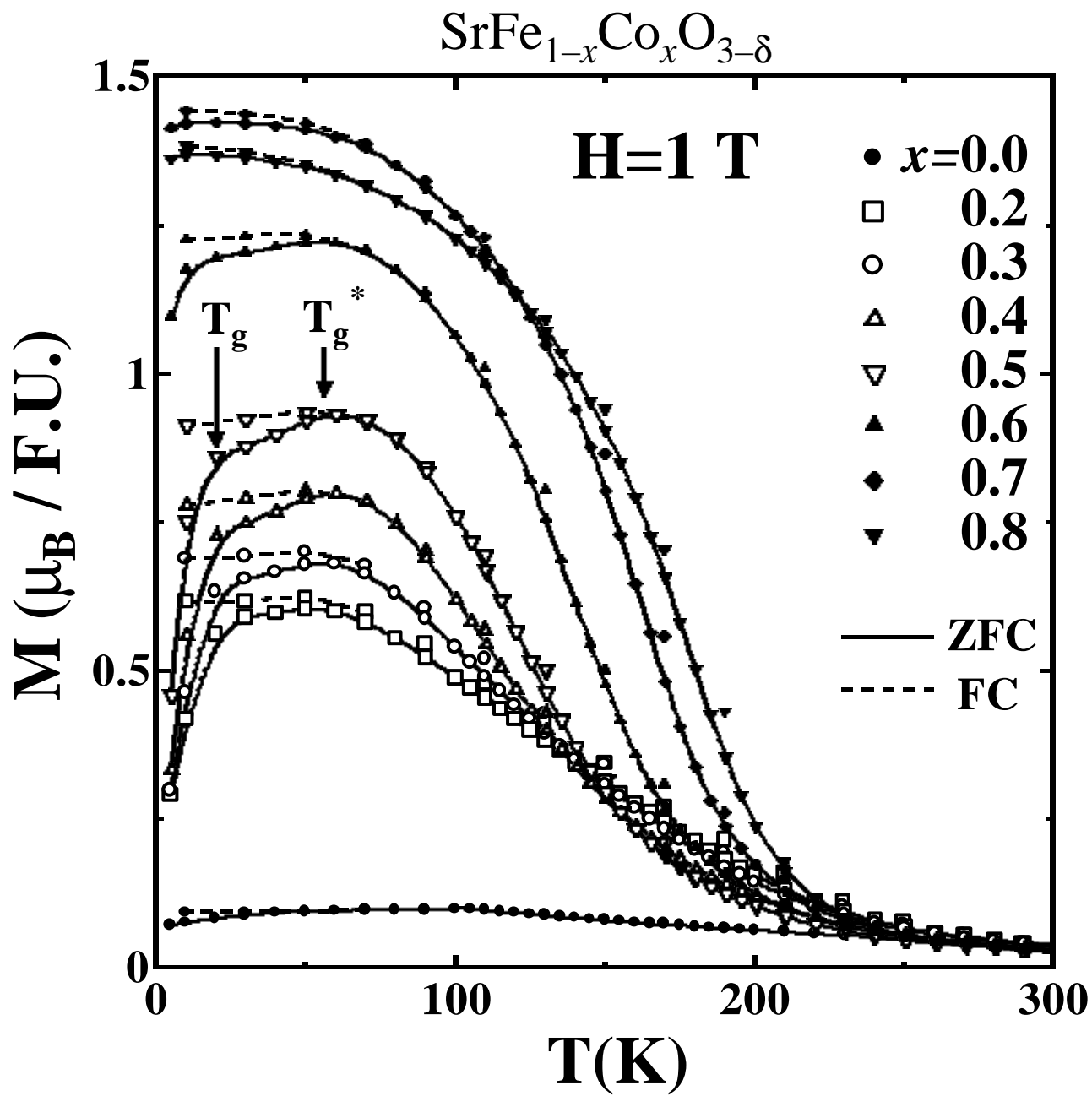

Fig. 3

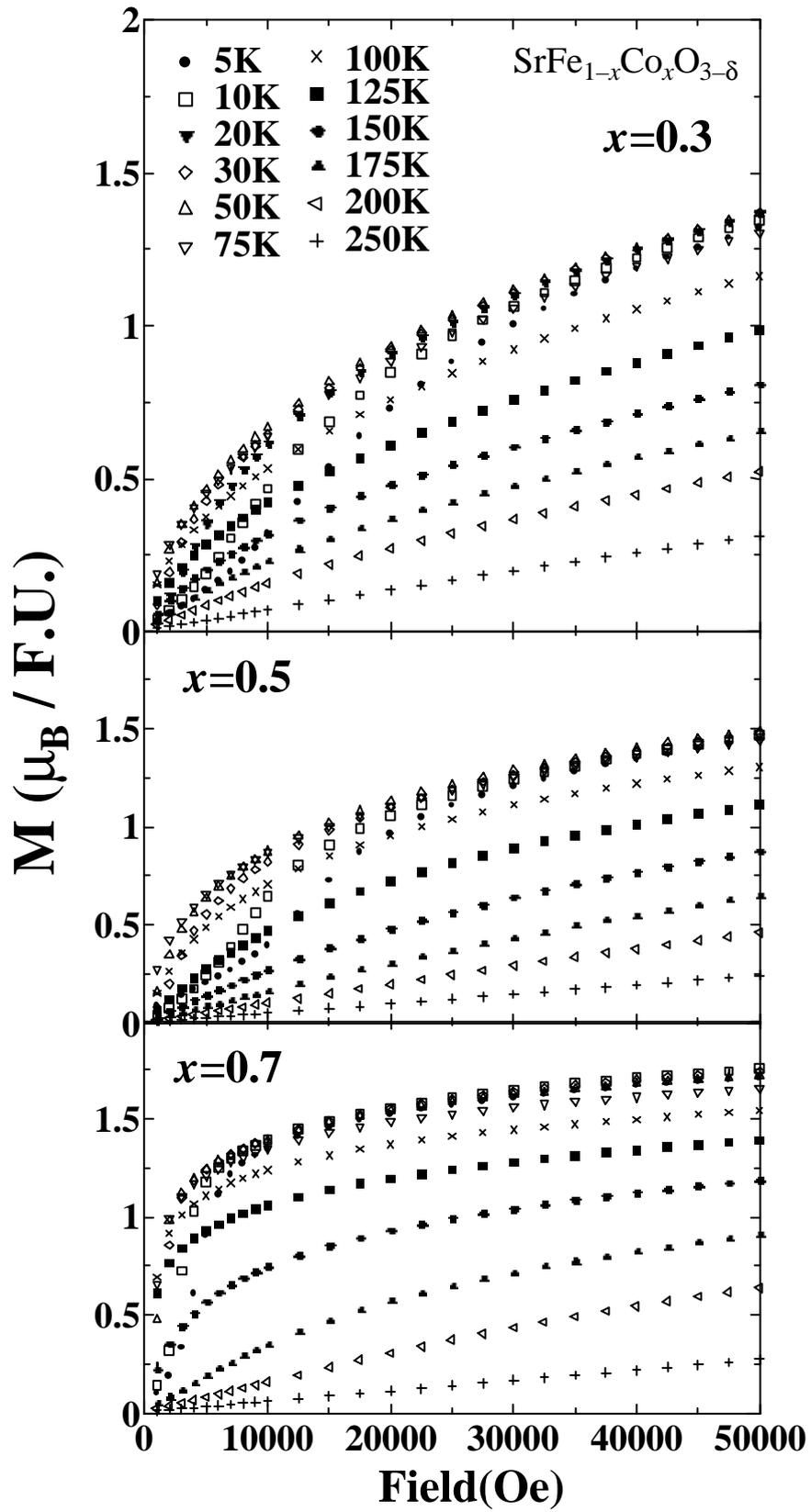

Fig. 4

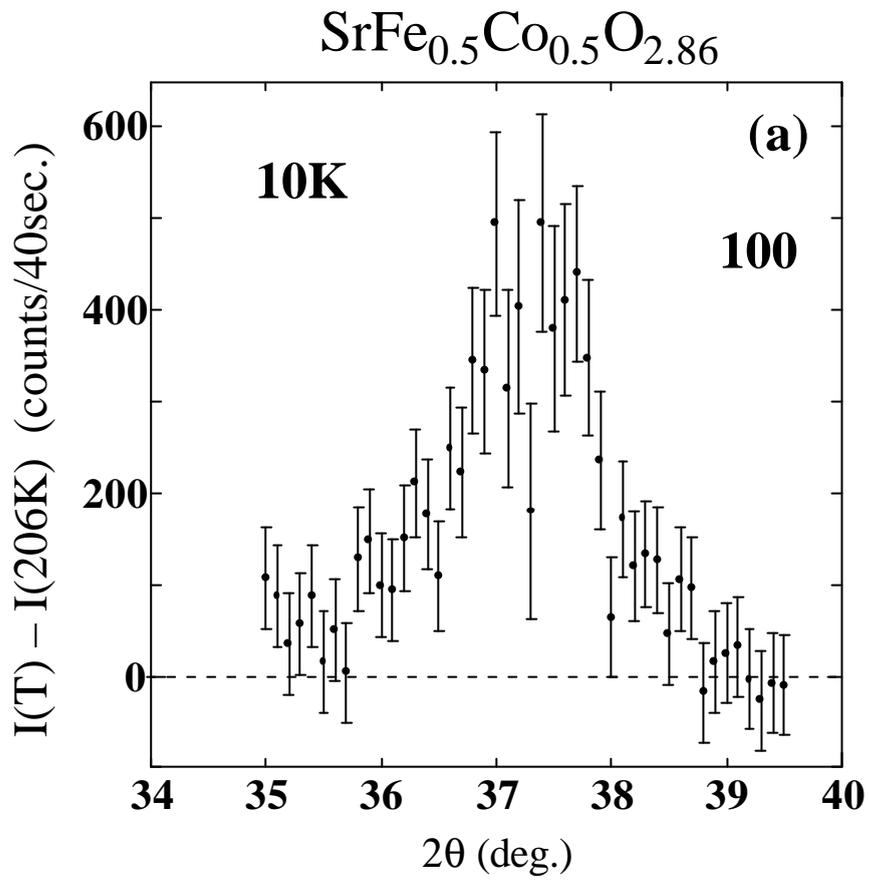

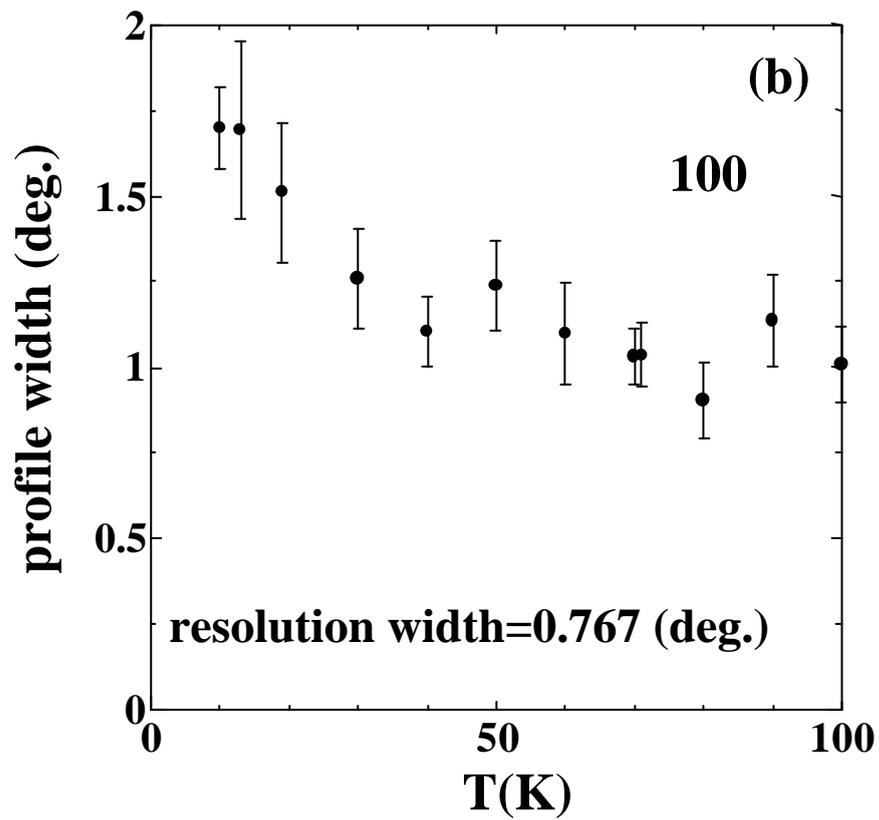

Fig. 5

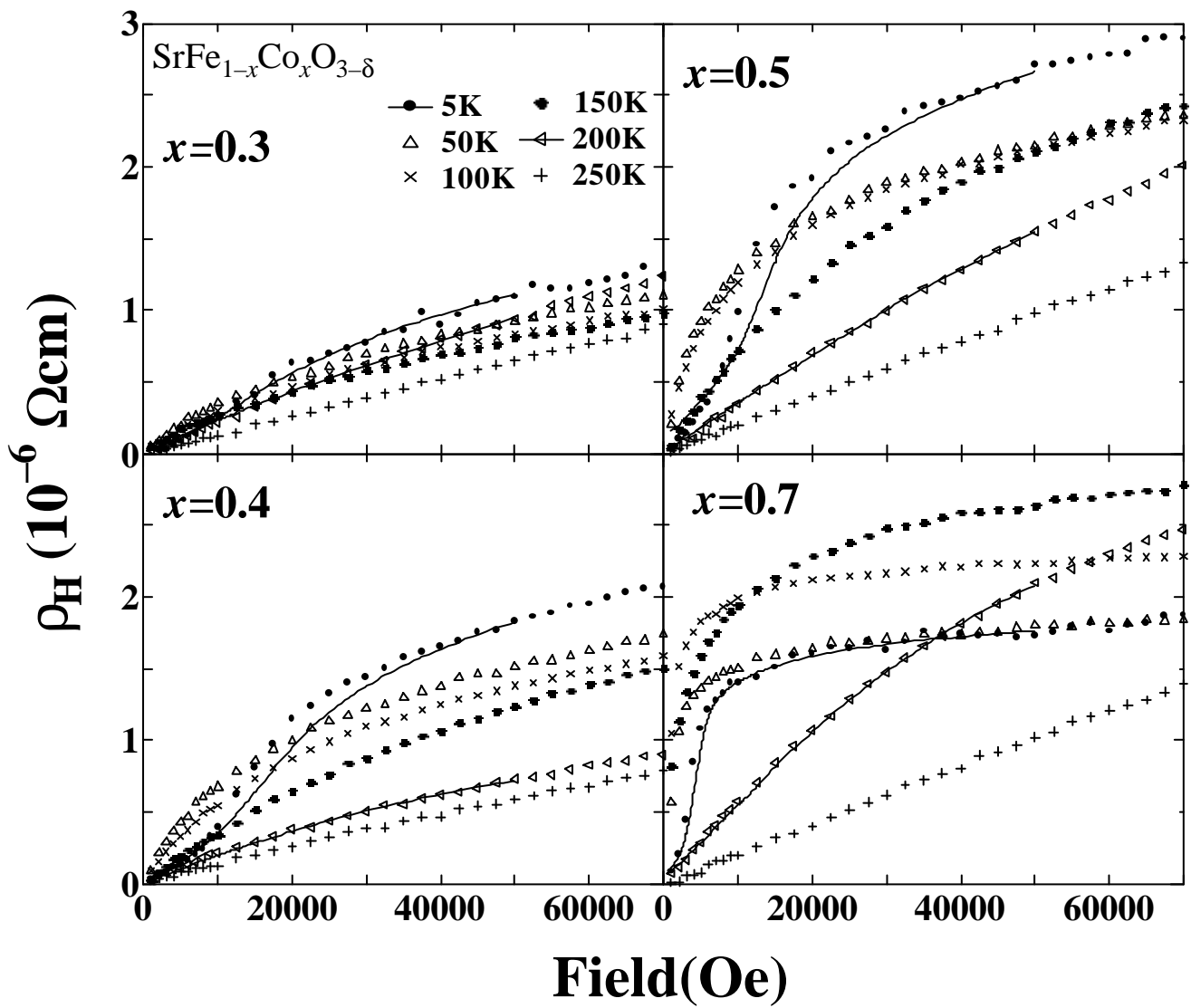

Fig. 6

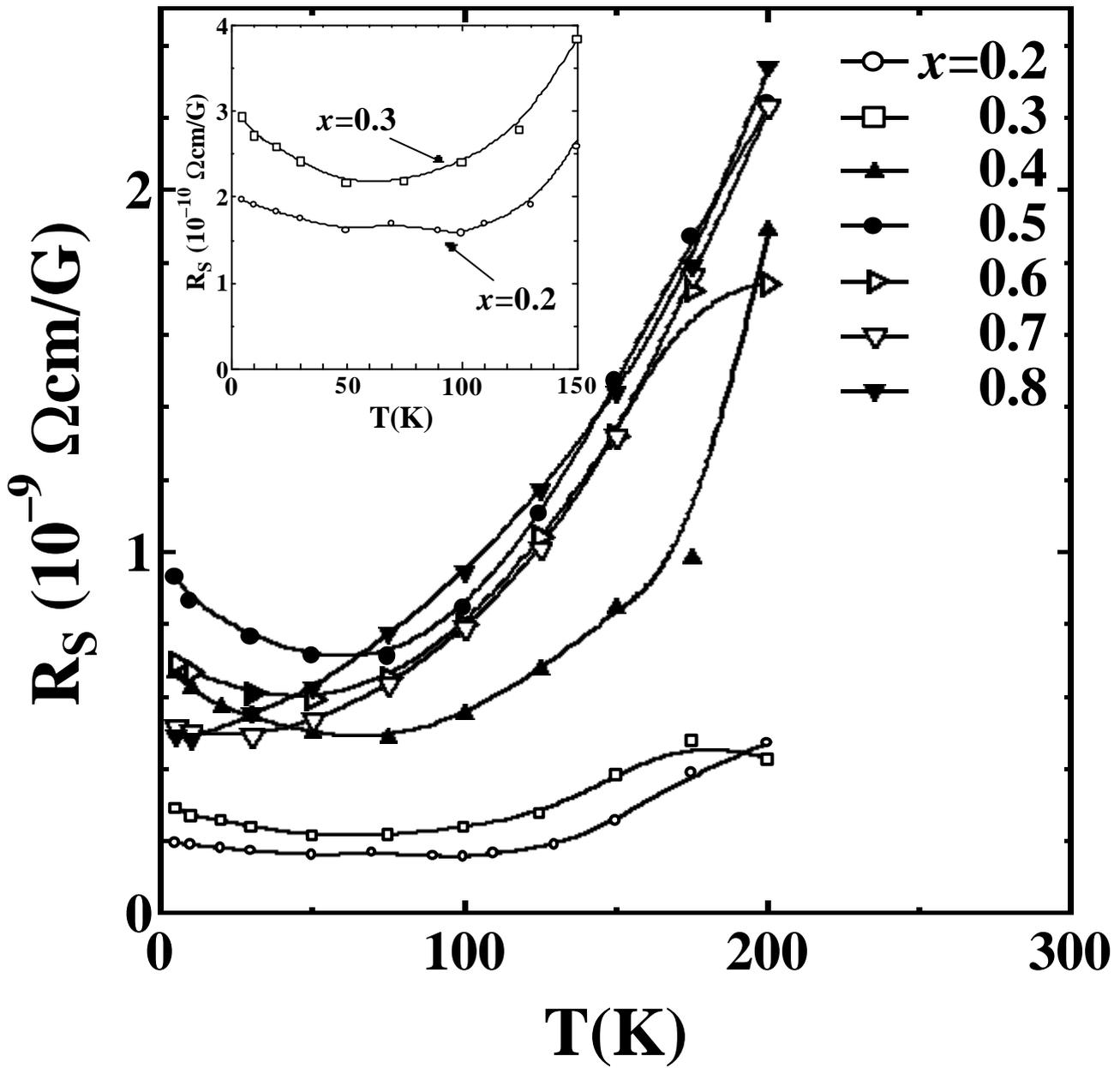

Fig. 7

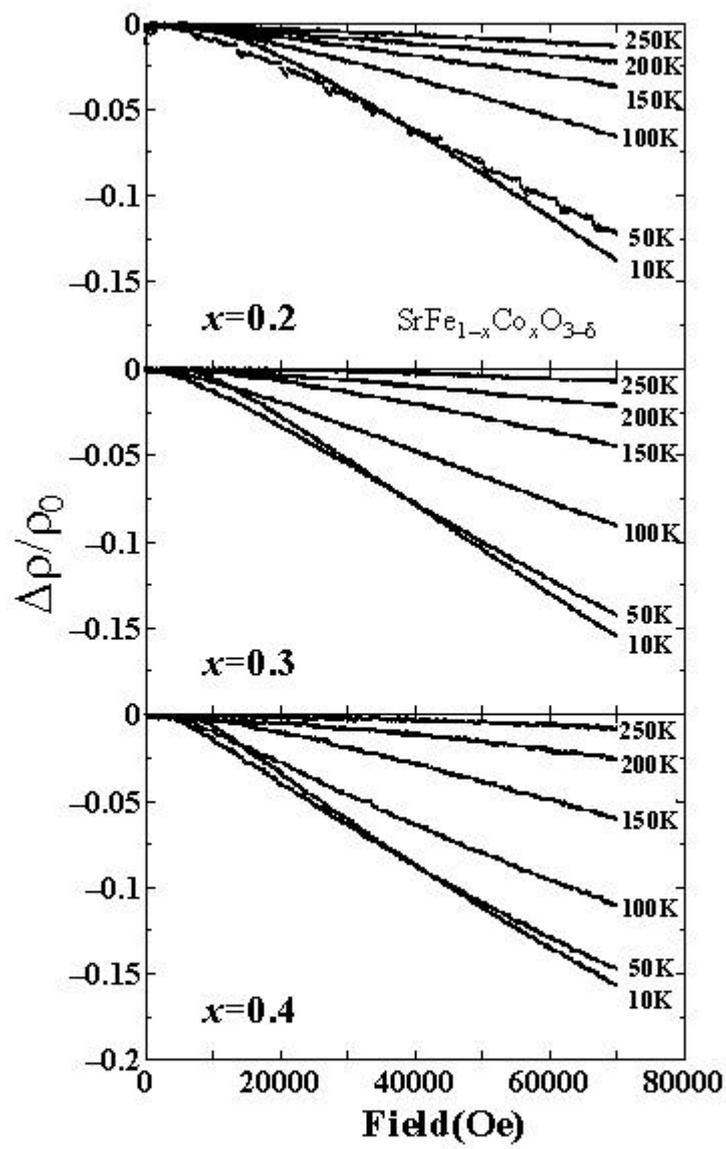

Fig. 8(a)

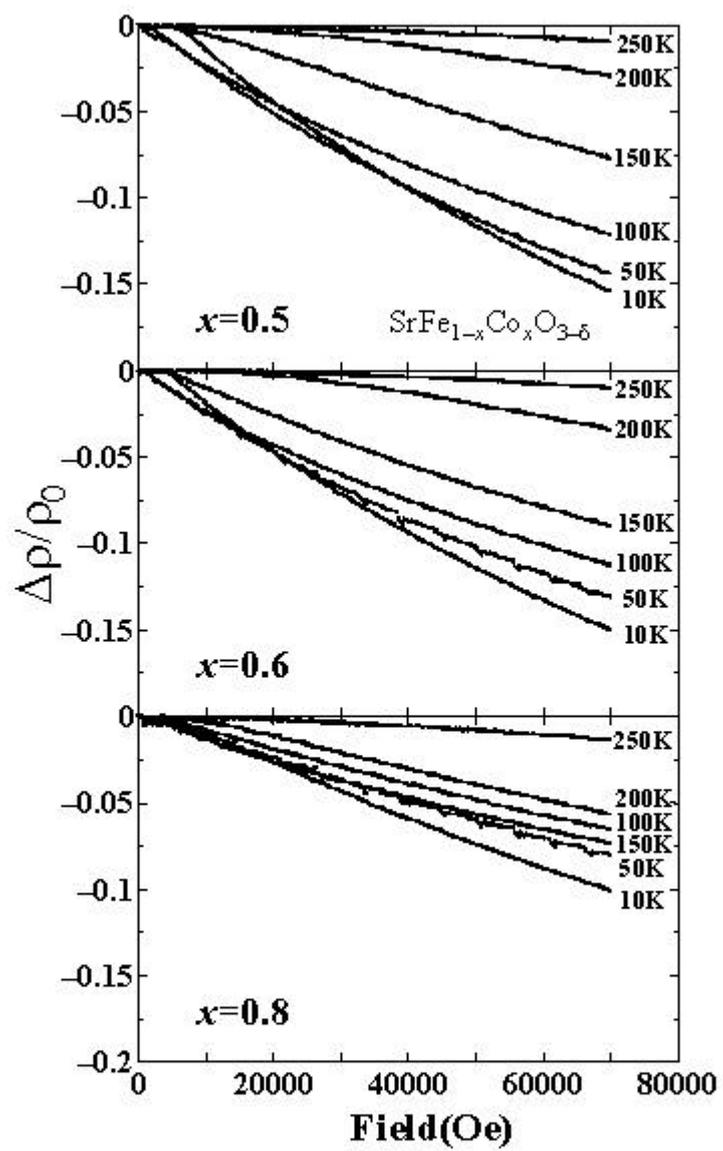

Fig. 8(b)

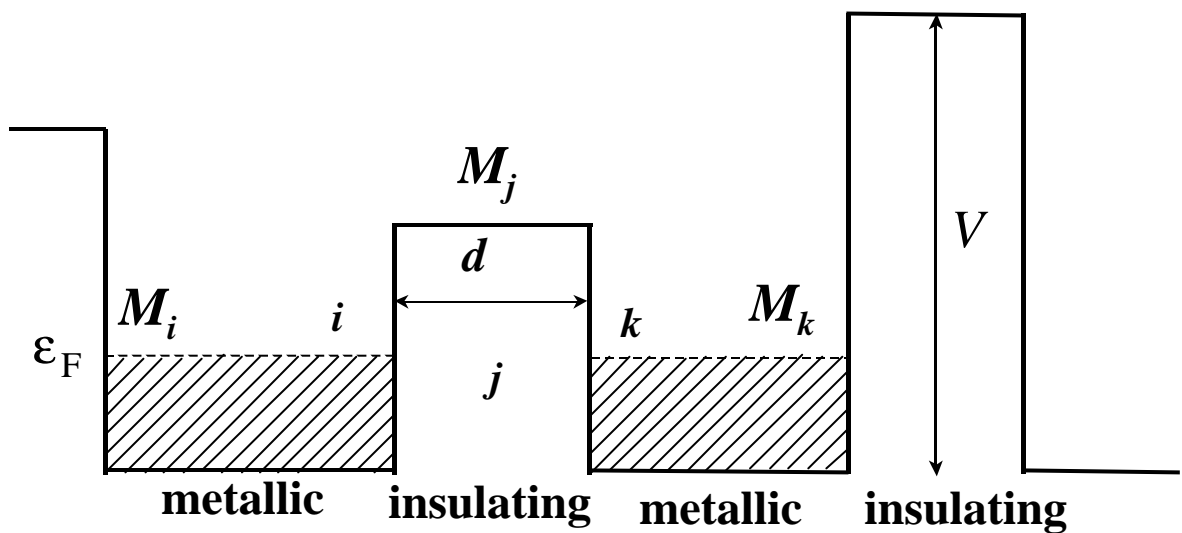

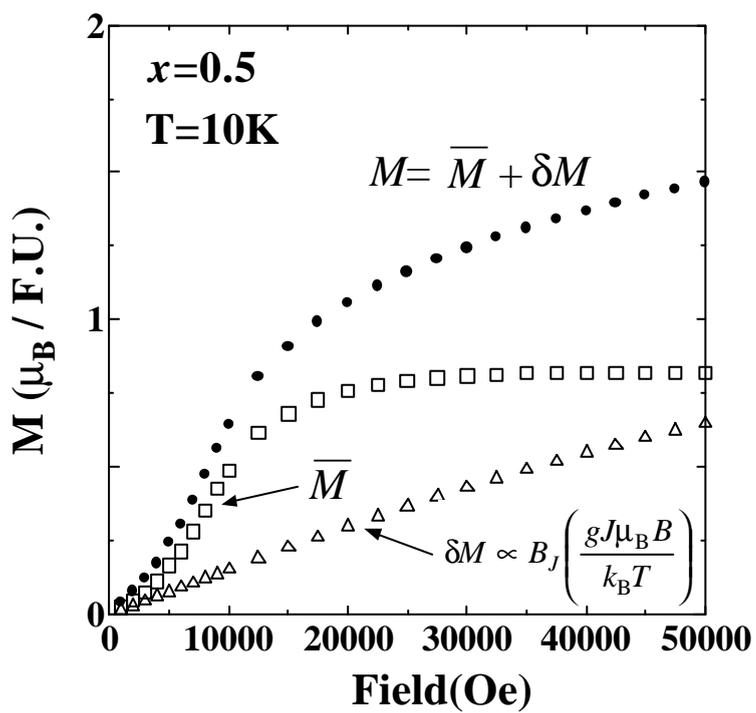

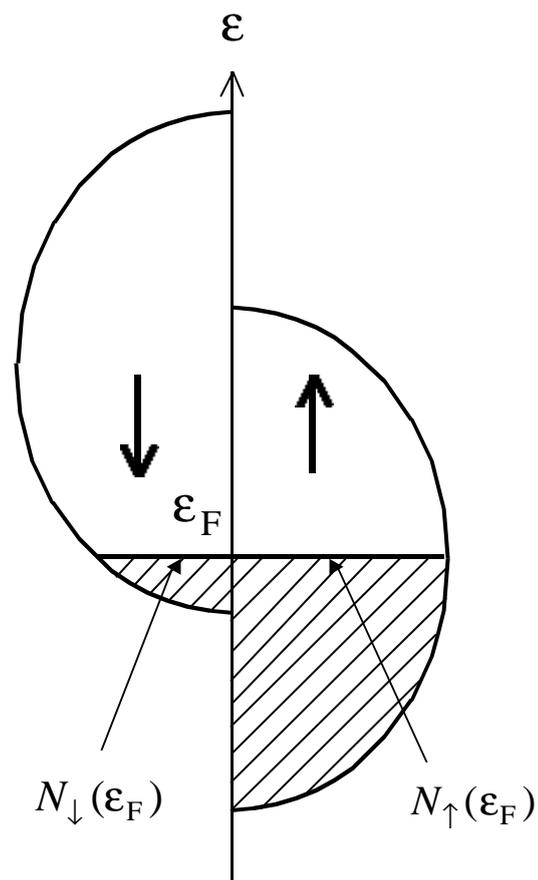

Fig. 9

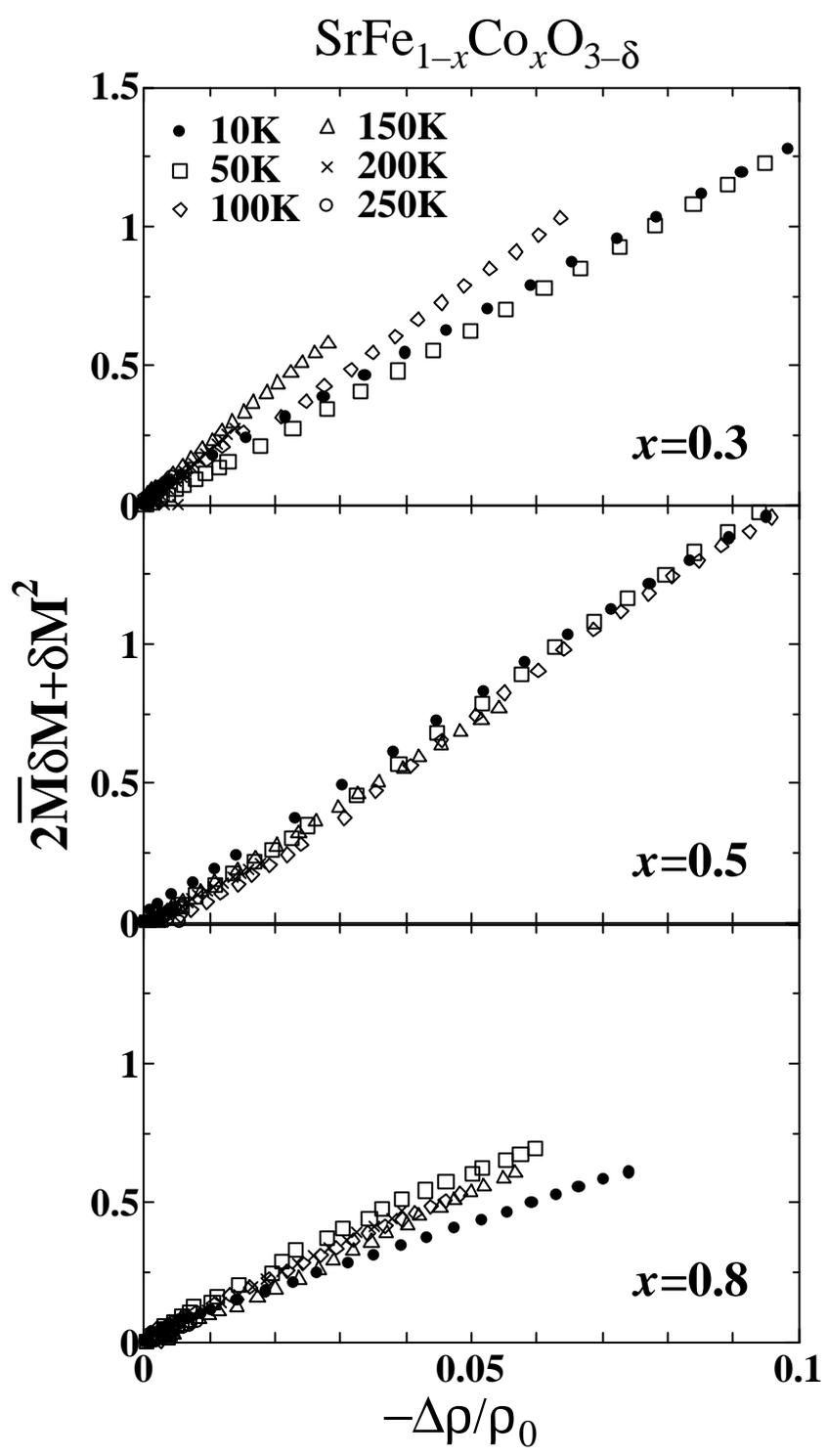

Fig. 10